\documentclass[pra,aps,twocolumn,amsmath,amssymb,showpacs]{revtex4}

\begin{document}
\clearpage
\preprint{}

\title{Uncertainty relations for arbitrary measurement in terms of
R\'{e}nyi entropies}

\author{Alexey E. Rastegin}
\affiliation{Department of Theoretical Physics, Irkutsk State University,
Gagarin Bv. 20, Irkutsk 664003, Russia}

\begin{abstract}
Uncertainty relations for a pair of arbitrary measurements and for a
single measurement are posed in the form of inequalities using the
R\'{e}nyi entropies. The formulation deals with discrete observables. 
Both the relations with state-dependent and state-independent
bounds are presented. The entropic bounds are illustrated within the 
distinction between non-orthogonal states.
\end{abstract}

\maketitle

\pagenumbering{arabic}
\setcounter{page}{1}

The Heisenberg uncertainty principle \cite{heisenberg} is first and
most known of those results that stress the primary features of the
quantum world. The progress of quantum theory has lead to a few
related insights such as the Bell inequality \cite{bell}, the quantum
Zeno effect \cite{misra}, the no-cloning theorem \cite{wootters1} and
the interaction-free measurement \cite{vaidman}. Although these
points stint our ability in manipulating quantum, they also clear
novel ways. For example, the Ekert scheme of quantum cryptography is
based on Bell's theorem \cite{ekert}. The techniques similar to
Zeno-effect behavior can be used for reducing decoherence in quantum
information processing \cite{viola}. The well-known quantitative form
of the uncertainty principle was given by Robertson \cite{robert}.
The standard deviations of observables ${\mathsf{A}}$ and ${\mathsf{B}}$ 
measured in the quantum state $|\psi\rangle$ satisfy
\begin{equation}
\Delta{\mathsf{A}}\>\Delta{\mathsf{B}}\geq
\frac{1}{2}\>|\langle\psi|[{\mathsf{A}},{\mathsf{B}}]|\psi\rangle|
\ . \label{rob}
\end{equation}
Due to the various scenarios of measurement, more specified relations
have been posed \cite{ozawa1,hall1,lahti}. For example, Bohr's
principle of complementarity \cite{bohr} was quantified by uncertainty 
relations. The most wide of these were obtained in Refs. \cite{hall1,ozawa2}.
In general, the key property of complementarity was studied by Kendon and
Sanders \cite{kendon}. Their approach follows the
information-theoretic viewpoint proposed by Wootters and Zurek
\cite{wootters2}.

The recent progress in the study of quantum information
processing shows that information theory gives a good base for
dealing with quantum properties. The concept of entropy is much
widely used in physics, both classical and quantum
\cite{wehrl,vedral}. So it is natural to state 
Heisenberg's principle in terms of entropies. The discussion of
entropic uncertainty relations with references will be
given below. Of late years, reformulations of the uncertainty 
principle have been posed, especially in information-theoretic terms:
in terms of the skew Wigner--Yanase information \cite{luo1}, the Fisher
information \cite{romera,luo2}, and the Holevo information \cite{winter1}. 
The uncertainty relation of Landau-Pollak type \cite{imai} and 
the one for two unitary operators \cite{spindel} should also be
cited. Such relations are beyond the scope of the present work.

Now both the entropic uncertainty relations and the R\'{e}nyi entropy
have been widely adopted in the researches of quantum systems. For
example, the entropic relations were applied to study of the
entanglement \cite{guhne}, the locking effect \cite{ballester} and
the special classes of observables \cite{wehner,vicente}. The topic
of paper by Bialynicki-Birula \cite{birula3} is most close to the
scope of the present work. He obtained the uncertainty relations in
terms of R\'{e}nyi entropies for the position--momentum and
angle--angular momentum pairs. The paper \cite{birula3} also contains
a list of references on fruitful applications of the R\'{e}nyi
entropy to quantum problems. Here we mention only recent papers
\cite{cubitt} in which the R\'{e}nyi entropy of quantum state is utilized.

The aim of the present work is to formulate the uncertainty
relations in terms of the R\'{e}nyi entropies for arbitrary
measurements. Important as the concrete observables are, they
is not able to give meaning of measurement limitations in all 
respects. The exposition is not restricted to von Neumann measurements
described by "Projector-Valued Measure" (PVM). We shall focus on
generalized ones described by "Positive Operator-Valued Measure"
(POVM). Recall that POVM $\{{\mathsf{M}}_i\}$ is a set of positive
operators ${\mathsf{M}}_i$ satisfying \cite{peres}
\begin{equation}
\sum\nolimits_i {\mathsf{M}}_i={\mathbf{1}}
\ , \label{povmdef}
\end{equation}
where ${\mathbf{1}}$ is the identity operator. This improved approach
to quantum measurements allows to extract more information from a
quantum system than von Neumann measurements \cite{peres}. The authors of
Ref. \cite{hillery} showed how to perform a generalized measurement via
a programmable quantum processor. There are a few ways to pose the uncertainty
principle for POVM measurement \cite{ozawa3}. In more recent paper by Massar
\cite{massar}, the uncertainty relations for POVM's are obtained by introducing
the uncertainty operator. He also examined the entropic uncertainty
relation in terms of the Shannon entropies for POVM's, whose elements
are all rank one \cite{massar}.

The first entropic relation was proposed by Hirschman \cite{hirs}. He
obtained the relation for position and momentum in terms of the
Shannon entropies. Hirschman also conjectured an improvement of his
result.  This conjecture has been proved by Beckner \cite{beck} and
by Bialynicki-Birula and Mycielski \cite{birula1}. A general proper
formulation of the uncertainty principle was asked by
Deutsch \cite{deutsch}. He emphasized that the right-hand side of Eq.
(\ref{rob}) is depended on $|\psi\rangle$. This leads to a
trivial lower bound for the deviations product, namely zero for any
eigenstate of either of operators ${\mathsf{A}}$ and ${\mathsf{B}}$
\cite{maass}. Deutsch obtained a lower bound on the sum of the
Shannon entropies for two observables without degeneracy. This formulation
has been completed by Partovi in some aspects \cite{hossein}. 
In particular, the case of degenerate observables has
been revealed. The position--momentum and angle--angular momentum pairs were
examined in Ref. \cite{hossein}. Improved bounds for the both pairs
were then obtained in Ref. \cite{birula2}.

It turned out that the entropic uncertainty relation given in Ref.
\cite{deutsch} can significantly be improved. The sharpened relation
has been conjectured by Kraus \cite{kraus} and then established by
Maassen and Uffink \cite{maass}. In Ref. \cite{hall2} Hall wrote that
the proof given by Maassen and Uffink directly extends to the case of
two POVM's, whose elements are all rank one. However, the formulation
stated in Ref. \cite{maass} deals with two non-degenerate
observables. A relevant extension for two degenerate observables
has been obtained by Krishna and Parthasarathy \cite{krishna}.
Using Naimark's theorem, they also got a lower bound on the sum of the
Shannon entropies for two generalized measurements. There is extension
of the result of Ref. \cite{maass} to more than two observables \cite{ruiz1}.
This result has also been sharpened in the cases of real \cite{gull} and complex
two-dimensional spaces \cite{ruiz2,ghirardi}.

Along with the Shannon entropy, other information entropies are
extensively used in the literature. One of them is so-called "Tsallis
entropy." This measure introduced by Havrda and Charv\'{a}t \cite{havrda}
became widely used in statistical mechanics after the fruitful work
of Tsallis \cite{tsallis}. The uncertainty relations in terms of the
Tsallis entropies were obtained for the position--momentum pair
\cite{raja} and the spin components \cite{majer}. Another generalization
of the Shannon entropy has been introduced by R\'{e}nyi \cite{renyi}. 
Larsen derived uncertainty relations in terms of the so-called purities
\cite{larsen}. The purity is immediately connected with the R\'{e}nyi 
entropy of order two.

We shall now describe the notation that is used throughout the text.
All logarithms are to the base two. Let $\alpha>0$ and
$\alpha\not=1$; then the R\'{e}nyi entropy of order $\alpha$ of
probability distribution $\{p_i\}$ is defined by \cite{renyi}
\begin{equation}
H_{\alpha}:=\frac{1}{1-\alpha}\> \log\left\{
\sum\nolimits_i p_i^{\alpha}
\right\} \ . \label{renent}
\end{equation}
This information measure is a nonincreasing function of parameter
$\alpha$; that is, if $\alpha<\beta$ then $H_{\alpha}\geq H_{\beta}$
\cite{renyi}. In the limit $\alpha\to1$ the R\'{e}nyi entropy tends
to the Shannon entropy $H_1\equiv-\sum\nolimits_i p_i\log p_i\>$.
For given measurement $\{{\mathsf{M}}_i\}$ and state $\rho$, the
probability of $i\,$th outcome is equal to 
$p_i={\rm tr}\{{\mathsf{M}}_i\rho\}$ \cite{peres}. The entropy
$H_{\alpha}({\mathsf{M}}|\rho)$ of generated probability distribution is
then defined by Eq. (\ref{renent}).

Let $\{{\mathsf{M}}_i\}$ and $\{{\mathsf{N}}_j\}$ be two POVM's. By definition,
we put the function
\begin{equation}
f({\mathsf{M}}{,}{\mathsf{N}}|\psi):=\underset{ij}{\max}
\>\frac{|\langle\psi|{\mathsf{M}}_i{\mathsf{N}}_j|\psi\rangle|}
{||{\mathsf{M}}_i^{1/2}|\psi\rangle||\>||{\mathsf{N}}_j^{1/2}|\psi\rangle||}
\ , \label{fmnpsi}
\end{equation}
where the maximum is taken over those values of labels $i$ and $j$ 
that the denominator is nonzero. In the case of mixed state $\rho$
with spectral decomposition
\begin{equation}
\rho=\sum\nolimits_{\lambda} \lambda\>|\psi_{\lambda}\rangle\langle\psi_{\lambda}|
\label{rhodec}
\end{equation}
we also define
\begin{equation}
f({\mathsf{M}}{,}{\mathsf{N}}|\rho):=\underset{\lambda}{\max}
\>f({\mathsf{M}}{,}{\mathsf{N}}|\psi_{\lambda})
\ . \label{fpqrho}
\end{equation}
For a single POVM $\{{\mathsf{M}}_i\}$ we put the function
\begin{equation}
\phi({\mathsf{M}}|\rho):=\underset{i}{\max}\>
{\rm tr}\{{\mathsf{M}}_i\rho\}
\ . \label{phidef}
\end{equation}
Below all the entropic bounds will be posed in terms of the defined
functions. The first result of the present work is stated as follows.

{\bf Relation 1} {\it For arbitrary two measurements
$\{{\mathsf{M}}_i\}$ and $\{{\mathsf{N}}_j\}$, there holds}
\begin{equation}
H_{\alpha}({\mathsf{M}}|\rho)+H_{\beta}({\mathsf{N}}|\rho)
\geq -2\log f({\mathsf{M}},{\mathsf{N}}|\rho)
 \ , \label{renrel1}
\end{equation}
{\it where orders $\alpha$ and $\beta$ satisfy
$1/\alpha+1/\beta=2$.}

In view of the previous result, this generalization seems 
to be plausible in itself. However, the proof of Relation 1 
requires a considerable alteration of the techniques in 
many respects. Writing out these tedious details, a perception
of the final results would be laboured. So the proof is given
in the comprehensive paper \cite{rast}, which presents a more 
complete account. It is valuable to get an entropic bound
for a single POVM. The ordinary way is to put
$\{{\mathsf{M}}_i\}=\{{\mathsf{N}}_j\}$ \cite{massar,krishna}.
But only for $\alpha\leq 1$ this way holds due to the 
nonincrease property \cite{birula3}. Meantime, the study 
of the R\'{e}nyi entropies of order two and three has
a clear physical motivation \cite{zycz}. In Ref. \cite{rast}
the following relation is obtained by other method.

{\bf Relation 2} {\it For arbitrary measurement $\{{\mathsf{M}}_i\}$
and each $\alpha>0$, there holds}
\begin{equation}
H_{\alpha}({\mathsf{M}}|\rho)
\geq -\log \phi({\mathsf{M}}|\rho)
\ . \label{renrel3}
\end{equation}

The lower bound (\ref{renrel1}) has been proved under the condition
$1/\alpha+1/\beta=2$. Bialynicki-Birula obtained the entropic
uncertainty relations for position and momentum under the same
condition \cite{birula3}. When orders $\alpha$ and $\beta$ are
not coupled in this way, we can write only
\begin{equation}
H_{\alpha}({\mathsf{M}}|\rho)+H_{\beta}({\mathsf{N}}|\rho)
\geq -\log\left[\phi({\mathsf{M}}|\rho)\phi({\mathsf{N}}|\rho)\right]
\ . \label{unrel1}
\end{equation}
This inequality obtained from Eq. (\ref{renrel3}) is valid for any two POVM's
and arbitrary $\alpha,\beta\in(0;+\infty)$. 

The entropic bounds presented above are dependent on state before measurement.
As it is already mentioned, in the Robertson relation (\ref{rob}) such a 
dependence leads to some unsuitability. In a sense, for the presented entropic 
relations this criticism may be refuted (for details, see \cite{rast}). Nevertheless,
we can at once get the state-independent bounds. It is easy to check \cite{rast} that
\begin{equation}
f({\mathsf{M}},{\mathsf{N}}|\rho)\leq\underset{ij}{\max}\>
||{\mathsf{M}}_i^{1/2}{\mathsf{N}}_j^{1/2}||
\ , \label{fmn}
\end{equation}
where the norm of operator ${\mathsf{Q}}$ is defined by
\begin{equation}
||{\mathsf{Q}}||:=\underset{\langle u|u\rangle=1}{\max}
||{\mathsf{Q}}\,|u\rangle||
\ . \label{normdef}
\end{equation}
Hence for arbitrary state $\rho$ we have
\begin{equation}
H_{\alpha}({\mathsf{M}}|\rho)+H_{\beta}({\mathsf{N}}|\rho)
\geq -2\log \underset{ij}{\max}\>
||{\mathsf{M}}_i^{1/2}{\mathsf{N}}_j^{1/2}||
 \ , \label{renrel1ind}
\end{equation}
where $1/\alpha+1/\beta=2$. Further, there holds \cite{rast}
\begin{equation}
H_{\alpha}({\mathsf{M}}|\rho)\geq
-\log\underset{i}{\max}\>||{\mathsf{M}}_i||
\ . \label{renrel3ind}	
\end{equation}
If we take in Eq. (\ref{renrel1ind}) the two POVM's to be identical, then
we just get Eq. (\ref{renrel3ind}). The right-hand side of Eq. 
(\ref{renrel3ind}) is an analogue of trivial zero bound for projective
measurement. Of course, Relation 2 is far more sharp.

Let us discuss an example of applications of the obtained relations.
Consider a game involving two parties, Alice and Bob. Alice
secretly chooses a state from the set
$\{|\psi_1\rangle,|\psi_2\rangle\}$ known to both parties.
She then sends the chosen state to Bob. His task is to identity the
state Alice has given him. Let us put
$|\pm\rangle\equiv(|0\rangle\pm|1\rangle)/\sqrt2$. In our simple case
we take $|\psi_1\rangle=|0\rangle$ and $|\psi_2\rangle=|+\rangle$.
Bob can use two different strategies. The strategy developed by
Helstrom \cite{helstrom} is not error-free. In our example,
Bob's optimal measurement is described by PVM
$\{{\mathsf{N}}_1,{\mathsf{N}}_2\}$ with elements
${\mathsf{N}}_{1}=|x\rangle\langle x|$ and
${\mathsf{N}}_{2}=|y\rangle\langle y|$, where
\begin{align}
|x\rangle &\equiv\cos(\pi/8)\>|0\rangle
-\sin(\pi/8)\>|1\rangle \ , \\
|y\rangle &\equiv\sin(\pi/8)\>|0\rangle
+\cos(\pi/8)\>|1\rangle \ .
\end{align}
If the outcome $N_1$ (the outcome $N_2$) is detected then Bob
concludes that Alice sent the state $|\psi_1\rangle$ (the state
$|\psi_2\rangle$). The above PVM minimizes the average probability of
mis-identification \cite{helstrom} equal to
$(1/2)\bigl\{\langle\psi_1|{\mathsf{N}}_{2}|\psi_1\rangle
+\langle\psi_2|{\mathsf{N}}_{1}|\psi_2\rangle\bigr\}
=(\sqrt2-1)\>2^{-3/2}$.

Bob can also use the unambiguous discrimination proposed by Ivanovich
\cite{ivan}, Dieks \cite{dieks} and Peres \cite{peres1}. If he allows 
the inconclusive answer then it is possible for him to perform a 
measurement without mis-identification. The optimal measurement 
minimizes the probability of inconclusive answer \cite{peres1}. Note 
that the unambiguous discrimination has important application to the quantum cryptography. 
In the B92 protocol \cite{bennett2}, Alice encode the bits 0 and 1 into
two nonorthogonal pure states. In Ref. \cite{bennett2} Bennett described 
the strategy, whose efficiency is less than 50 $\%$. The authors of 
Ref. \cite{palma} built the procedure based on the unambiguous discrimination.
The efficiency of proposed strategy is greater than 50 $\%$. 

Let us return to the discussed game. We consider a POVM
$\{{\mathsf{M}}_1,{\mathsf{M}}_2,{\mathsf{M}}_3\}$ with elements
\cite{peres1}
\begin{align}
{\mathsf{M}}_1 &=\sqrt2\>(\sqrt2+1)^{-1}
\>|-\rangle\langle-| \ , \\
{\mathsf{M}}_2 &=\sqrt2\>(\sqrt2+1)^{-1}
\>|1\rangle\langle1| \ , \\
{\mathsf{M}}_3 &={\mathbf{1}}-{\mathsf{M}}_1-{\mathsf{M}}_2 \ .
\end{align}
If the outcome $M_1$ (the outcome $M_2$) is detected then Bob can
exactly conclude that the state $|\psi_1\rangle$ (the state
$|\psi_2\rangle$) has been received. Sometimes, however, the
inconclusive outcome $M_3$ will occur, and then he will obtain no
information about the received state. The probability of this answer
is $|\langle\psi_1|\psi_2\rangle|=1/\sqrt2$ \cite{peres1}.

The discussed two POVM's $\{{\mathsf{M}}_i\}$ and $\{{\mathsf{N}}_j\}$
have only one-rank elements. It turns out that in this case the inequality
(\ref{fmn}) is saturated regardless of state $\rho$ \cite{rast}.
So, $f({\mathsf{M}},{\mathsf{N}}|\rho)^2=||{\mathsf{M}}_1^{1/2}{\mathsf{N}}_1||^2
=||{\mathsf{M}}_2^{1/2}{\mathsf{N}}_2||^2=1/2$ 
by calculations. Relation 1 then leads to
\begin{equation}
H_{\alpha}({\mathsf{M}}|\rho)+H_{\beta}({\mathsf{N}}|\rho)\geq 1
\label{exam1}
\end{equation}
for any state $\rho$. On the contrary, the right-hand side of Eq. (\ref{unrel1})
is significantly depend on a quantum state. Let us consider this inequality
for the state $|\psi_1\rangle=|0\rangle$. By calculations, we have
\begin{align}
& \phi({\mathsf{M}}|\psi_1)=\langle 0|{\mathsf{M}}_3|0\rangle
=2^{-1/2} \ , \label{mpsi1} \\	
& \phi({\mathsf{N}}|\psi_1)=\langle 0|{\mathsf{N}}_1|0\rangle
=2^{-3/2}(\sqrt{2}+1) \ . \label{npsi1} 	
\end{align}
The entropic relation (\ref{unrel1}) then gives
\begin{equation}
H_{\alpha}({\mathsf{M}}|\psi_1)+H_{\beta}({\mathsf{N}}|\psi_1)
\geq 2-\log\,(\sqrt2+1)
\approx 0.728
\ . \label{exam2} 
\end{equation}
The inequality (\ref{exam1}) is stronger than the inequality (\ref{exam2}).
The difference 0.272 between the right-hand sides of Eq. (\ref{exam1})
and Eq. (\ref{exam2}) may be regarded as a manifestation of incompatibility
of the two discussed measurements. We shall now compare the entropic bounds in 
Eqs. (\ref{renrel3}) and (\ref{renrel3ind}) within our example. For the state 
$|\psi_1\rangle$ Relation 2 gives $H_{\alpha}({\mathsf{M}}|\psi_1)\geq 0.5$
according to Eq. (\ref{mpsi1}). This considerably exceeds the trivial lower bound 
$-\log ||{\mathsf{M}}_3||=\log\,(\sqrt2+1)-1\approx 0.272$ given
by Eq. (\ref{renrel3ind}). We see that the state-dependent bound in Eq. 
(\ref{renrel3}) can be far stronger. Thus, Relation 2 ensures a nontrivial 
entropic bound for a single POVM.

In our example we have seen that the lower bound (\ref{renrel1}) can be
more sharp than the lower bound (\ref{unrel1}). As it is shown in Ref. \cite{rast},
in one's turn the lower bound (\ref{unrel1}) can be more sharp than the lower bound 
(\ref{renrel1}). In addition, the correctness of Eq. (\ref{unrel1}) is not 
limited by the condition $1/\alpha+1/\beta=2$. So both the bounds (\ref{renrel1})
and (\ref{unrel1}) should be considered as independent entropic
relations. It was above mentioned that the bound given by Maassen and
Uffink \cite{maass} has been sharpened in some cases. So there is
clear scope for improvement of the lower bounds (\ref{renrel1}) and
(\ref{unrel1}). These questions remain for the future.

{\bf Acknowledgments.} I am grateful to an anonymous referee for helpful remarks.


\begin{thebibliography}{99}

\bibitem{heisenberg}%------------------------------------------------
W.~Heisenberg, Z. Phys. {\bf 43}, 172 (1927)

\bibitem{bell}%------------------------------------------------
J.~S.~Bell, Physics {\bf 1}, 195 (1964)

\bibitem{misra}%------------------------------------------------
B.~Misra and E.~C.~G.~Sudarshan, J. Math. Phys. {\bf 18}, 756 (1977)

\bibitem{wootters1}%-----------------------------------------------
W.~K.~Wootters and W.~H.~Zurek, Nature (London) {\bf 299}, 802 (1982)

\bibitem{vaidman}%----------------------------------------------
A.~C.~Elitzur and L.~Vaidman, Found. Phys. {\bf 23}, 987 (1993)

\bibitem{ekert}%----------------------------------------------
A.~K.~Ekert, Phys. Rev. Lett. {\bf 67}, 661 (1991)

\bibitem{viola}%----------------------------------------------------
L.~Viola and S.~Lloyd, Phys. Rev. A {\bf 58}, 2733 (1998)

\bibitem{robert}%--------------------------------------------------
H.~P.~Robertson, Phys. Rev. {\bf 34}, 163 (1929)

\bibitem{ozawa1}%-----------------------------------------------
M.~Ozawa, Phys. Rev. A {\bf 67}, 042105 (2003)

\bibitem{hall1}%-----------------------------------------------
M.~J.~W.~Hall, Phys. Rev. A {\bf 69}, 052113 (2004)

\bibitem{lahti}%-----------------------------------------------
P.~Busch, T.~Heinonen, and P.~J.~Lahti, Phys. Rep. {\bf 452}, 155
(2007)

\bibitem{bohr}%-------------------------------------------------
N.~Bohr, Nature (London) {\bf 121}, 580 (1928)

\bibitem{ozawa2}%-----------------------------------------------
M.~Ozawa, Phys. Lett. A {\bf 320}, 367 (2004)

\bibitem{kendon}%-----------------------------------------------------
V.~Kendon and B.~C.~Sanders, Phys. Rev. A {\bf 71}, 022307 (2005)

\bibitem{wootters2}%-----------------------------------------------
W.~K.~Wootters and W.~H.~Zurek, Phys. Rev. D {\bf 19}, 473 (1979)

\bibitem{wehrl}%-----------------------------------------------
A.~Wehrl, Rev. Mod. Phys. {\bf 50}, 221 (1978)

\bibitem{vedral}%-----------------------------------------------
V.~Vedral, Rev. Mod. Phys. {\bf 74}, 197 (2002)

\bibitem{luo1}%-------------------------------------------------
S.~Luo, Phys. Rev. Lett. {\bf 91}, 180403 (2003); S.~Luo, Phys.
Rev. A {\bf 72}, 042110 (2005)

\bibitem{luo2}%-------------------------------------------------
S.~Luo, Lett. Math. Phys. {\bf 53}, 243 (2000); S.~Luo and Z.~Zhang,
J. Stat. Phys. {\bf 114}, 1557 (2004); P.~Gibilisco, D.~Imparato, and
T.~Isola, J. Stat. Phys. {\bf 130}, 545 (2008)

\bibitem{romera}%-------------------------------------------------
E.~Romera, P.~S\'anchez-Moreno, and J.~S.~Dehesa, J. Math. Phys.
{\bf 47}, 103504 (2006); P.~S\'anchez-Moreno, R.~Gonz\'alez-F\'erez,
and J.~S.~Dehesa, New J. Phys. {\bf 8}, 330 (2006)

\bibitem{winter1}%-------------------------------------------------
M.~Christandl and A.~Winter, IEEE Trans. Inf. Theory {\bf 51}, 3159
(2005)

\bibitem{imai}%-------------------------------------------------
T.~Miyadera and H.~Imai, Phys. Rev. A {\bf 76}, 062108 (2007)

\bibitem{spindel}%-------------------------------------------------
S.~Massar and P.~Spindel, Phys. Rev. Lett. {\bf 100}, 190401 (2008)

\bibitem{guhne}%--------------------------------------------------
V.~Giovannetti, Phys. Rev. A {\bf 70}, 012102 (2004); O.~G\"{u}hne
and M.~Lewenstein, Phys. Rev. A {\bf 70}, 022316 (2004)

\bibitem{ballester}%----------------------------------------------
M.~A.~Ballester and S.~Wehner, Phys. Rev. A {\bf 75}, 022319 (2007)

\bibitem{wehner}%---------------------------------------------
S.~Wehner and A.~Winter, arXiv: 0710.1185 [quant-ph]

\bibitem{vicente}%----------------------------------------------
J.~I. de~Vicente and J.~S\'anchez-Ruiz, Phys. Rev. A {\bf 77}, 042110 (2008)

\bibitem{birula3}%-------------------------------------------------
I.~Bialynicki-Birula, Phys. Rev. A {\bf 74}, 052101 (2006)

\bibitem{cubitt}%-------------------------------------------------
T.~Cubitt, A.~W.~Harrow, D.~Leung, A.~Montanaro, and A.~Winter,
arXiv: 0712.3628 [quant-ph]; N.~Schuch, M.~M.~Wolf, F.~Verstraete, 
and J.~I.~Cirac, Phys. Rev. Lett. {\bf 100}, 030504 (2008);
R.~Augusiak, J.~Stasi\'nska, and P.~Horodecki, Phys. Rev. A {\bf 77},
012333 (2008); B.~Dierckx, M.~Fannes, and C.~Vandenplas, Phys. Rev. 
A {\bf 77}, 060302(R) (2008)

\bibitem{peres}%-----------------------------------------------
A.~Peres, {\it Quantum Theory: Concepts and Methods} (Kluwer,
Dordrecht, 1995)

\bibitem{hillery}%-----------------------------------------------
M.~Ro\v{s}ko, V.~Bu\v{z}ek, P.~R.~Chouha, and M.~Hillery, Phys. Rev. A
{\bf 68}, 062302 (2003)

\bibitem{ozawa3}%-----------------------------------------------
M.~Ozawa, Ann. Phys. (N.Y.) {\bf 311}, 350 (2004)

\bibitem{massar}%-----------------------------------------------
S.~Massar, Phys. Rev. A {\bf 76}, 042114 (2007)

\bibitem{hirs}%-------------------------------------------------
I.~I.~Hirschman, Am. J. Math. {\bf 79}, 152 (1957)

\bibitem{beck}%-------------------------------------------------
W.~Beckner, Ann. Math. {\bf 102}, 159 (1975)

\bibitem{birula1}%-------------------------------------------------
I.~Bialynicki-Birula and J.~Mycielski, Commun. Math. Phys. {\bf 44},
129 (1975)

\bibitem{deutsch}%-----------------------------------------------------
D.~Deutsch, Phys. Rev. Lett. {\bf 50}, 631 (1983)

\bibitem{maass}%-----------------------------------------------------
H.~Maassen and J.~B.~M.~Uffink, Phys. Rev. Lett. {\bf 60}, 1103 (1988)

\bibitem{hossein}%---------------------------------------------------
M.~H.~Partovi, Phys. Rev. Lett. {\bf 50}, 1883 (1983)

\bibitem{birula2}%-------------------------------------------------
I.~Bialynicki-Birula, Phys. Lett. A {\bf 103}, 253 (1984)

\bibitem{kraus}%---------------------------------------------------
K.~Kraus, Phys. Rev. D {\bf 35}, 3070 (1987)

\bibitem{hall2}%-----------------------------------------------
M.~J.~W.~Hall, Phys. Rev. A {\bf 55}, 100 (1997)

\bibitem{krishna}%--------------------------------------------------
M.~Krishna and K.~R.~Parthasarathy, Sankhya, Ser. A {\bf 64}, 842
(2002)

\bibitem{ruiz1}%----------------------------------------------------
J.~S\'anchez, Phys. Lett. A {\bf 173}, 233 (1993)

\bibitem{gull}%-----------------------------------------------------
A.~J.~M.~Garrett and S.~F.~Gull, Phys. Lett. A {\bf 151}, 453 (1990)

\bibitem{ruiz2}%-------------------------------------------------
J.~S\'anchez-Ruiz, Phys. Lett. A {\bf 244}, 189 (1998)

\bibitem{ghirardi}%----------------------------------------------------
G.-C.~Ghirardi, L.~Marinatto, and R.~Romano, Phys. Lett. A {\bf 317},
32 (2003)

\bibitem{havrda}%-------------------------------------------------
J.~Havrda and F.~Charv\'{a}t, Kybernetika {\bf 3}, 30 (1967)

\bibitem{tsallis}%-------------------------------------------------
C.~Tsallis, J. Stat. Phys. {\bf 52}, 479 (1988)

\bibitem{raja}%-------------------------------------------------
A.~K.~Rajagopal, Phys. Lett. A {\bf 205}, 32 (1995)

\bibitem{majer}%-------------------------------------------------
V.~Majernik and E.~Majernikova, Rep. Math. Phys. {\bf 47}, 381 (2001)

\bibitem{renyi}%-------------------------------------------------
A.~R\'{e}nyi, {\it On Measures of Entropy and Information},
Proc. Fourth Berkeley Symp. Math. Stat. Prob. (Vol. 1., University of
California Press, Berkeley, 1961), p. 547

\bibitem{larsen}%-----------------------------------------------
U.~Larsen, J. Phys. A: Math. Gen. {\bf 23}, 1041 (1990) 

\bibitem{rast}%----------------------------------------------------
A.~E.~Rastegin, arXiv: 0807.2691 [quant-ph]

\bibitem{zycz}%----------------------------------------------------
K.~\.{Z}yczkowski, Open Sys. Inf. Dyn. {\bf 10}, 297 (2003)

\bibitem{helstrom}%--------------------------------------------
C.~W.~Helstrom, {\it Quantum Detection and Estimation Theory}
(Academic Press, New York, 1976)

\bibitem{ivan}%-----------------------------------------------
I.~D.~Ivanovic, Phys. Lett. A {\bf 123}, 257 (1987)

\bibitem{dieks}%-----------------------------------------------
D.~Dieks, Phys. Lett. A {\bf 126}, 303 (1988)

\bibitem{peres1}%-----------------------------------------------
A.~Peres, Phys. Lett. A {\bf 128}, 19 (1988)

\bibitem{bennett2}%----------------------------------------------
C.~H.~Bennett, Phys. Rev. Lett. {\bf 68}, 3121 (1992)

\bibitem{palma}%----------------------------------------------
A.~K.~Ekert, B.~Huttner, G.~M.~Palma, and A.~Peres, Phys. Rev. A
{\bf 50}, 1047 (1994)

\end{thebibliography}
\end{document}